\documentclass[aps,prb,twocolumn,superscriptaddress]{revtex4}%
\usepackage{graphicx}
\usepackage{amsmath}
\usepackage{amsfonts}
\usepackage{amssymb}%
\providecommand{\U}[1]{\protect\rule{.1in}{.1in}}

\newcommand{\Lor}{\mathcal{L}_0}
\newcommand{\tempnote}[1]   {\begingroup{\it (NOTE: #1)}\endgroup}
\renewcommand{\tempnote}[1]{}
\begin{document}
\title{Spin heat accumulation and its relaxation in spin valves}
\author{T.~T. Heikkil\"a}
\email{Tero.Heikkila@tkk.fi}
\affiliation{Low Temperature Laboratory, Aalto University, P.O. Box 15100, FI-00076 AALTO, Finland}
\author{Moosa Hatami}
\author{Gerrit E. W. Bauer}
\affiliation{Kavli Institute of NanoScience, Delft University of Technology, 2628 CJ Delft,
The Netherlands}



\date{\today}

\begin{abstract}
We study the concept of spin heat accumulation in excited spin
valves, more precisely the effective electron temperature that may
become spin dependent, both in linear response and far from
equilibrium. A temperature or voltage gradient create
non-equilibrium energy distributions of the two spin ensembles in
the normal metal spacer, which approach Fermi-Dirac functions
through energy relaxation mediated by electron-electron and
electron-phonon coupling. Both mechanisms also exchange energy
between the spin subsystems. This inter-spin energy exchange may
strongly affect thermoelectric properties spin valves, leading,
\textit{e.g}., to violations of the Wiedemann-Franz law.

\end{abstract}
\maketitle

The electric conductance through ferromagnet$|$normal
metal$|$ferromagnet spin valves is a function of the magnetic
configuration.\cite{gmrref} It reflects the spin accumulation,
\textit{i.e.}, the spin (index $\sigma$) dependent chemical
potential $\mu_{\sigma}$ of the normal-metal island. The latter
parameterizes the spin dependence of the energy distribution
functions $f_{\sigma}(E)$, whose description also requires
spin-dependent temperatures $T_{\sigma}$.\cite{giazotto05,hatami}
As shown below, these should in general be interpreted as effective
parameters.

In this Rapid Communication we describe the processes affecting the $T_{\sigma}$
and through them the thermoelectric response in spin valves, which
we find to be a sensitive probe for the non-equilibrium state in the
non-magnetic spacer. Whereas the spin accumulation relaxes only by
scattering processes that break spin rotation invariance such as
spin-orbit interaction and magnetic disorder, the spin heat
accumulation $T_{s}=T_{\uparrow}-T_{\downarrow}$ is sensitive also
to electron-phonon (e-ph) and electron-electron (e-e) interactions.
Spin-flip scattering in Al, Ag, Cu, or carbon is weak and hardly
temperature dependent; the typical spin-flip scattering time
$\tau_{\mathrm{sf}}$ is of the order 100
$%
\operatorname{ps}%
$,\cite{materials} which can be much longer than the dwell times in
magnetoelectronic structures. The inter-spin energy exchange rate
due to inelastic effects is strongly temperature dependent and above
cryogenic temperatures typically dominates the direct spin-flip
scattering in dissipating the spin heat accumulation. The spin heat
accumulation in normal metal spacers should not be confused with the
spin (wave) temperature of ferromagnets.\cite{beaurepaire96}

\begin{figure}[h]
\centering
\includegraphics[width=0.71\columnwidth]{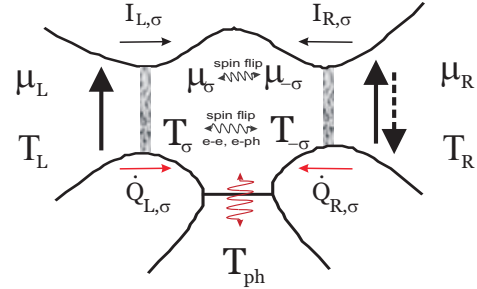}\caption{(Color online):
Schematic spin-valve biased with a voltage and/or temperature
difference. Spin-flip and inelastic electron-electron and
electron-phonon scattering in the normal metal spacer lead to
inter-spin energy exchange and change the thermoelectric
characteristics. $I$ and $\dot{Q}$ stand for the charge and heat
currents flowing into the island. $T_{\mathrm{ph}}$ is the
temperature of
the phonon bath. }%
\label{fig:spinvalve}%
\end{figure}

In a spin valve (Fig.~\ref{fig:spinvalve}), a nonmagnetic island is
coupled to two ferromagnetic reservoirs with parallel (P) or
antiparallel (AP) magnetic alignments. The chemical potential of the
left (right) reservoir is $\mu_{L(R)}$ and the temperature is
$T_{L(R)}$. The conductances $G_{L/R\sigma}$ and Seebeck
coefficients $S_{L/R\sigma}$ of the contacts between the island and
the reservoirs depend on spin $\sigma \in\{\uparrow,\downarrow\}$.
Biasing the spin valve with either a voltage $\Delta
V=(\mu_{R}-\mu_{L})/e$ or a temperature difference $\Delta
T=T_{R}-T_{L}$ gives rise to a spin-dependent energy distribution
function $f_{\sigma}(E)$ of the electrons on the island. As shown
below, in the linear response regime this can be described exactly
by spin-dependent chemical potentials and temperatures, such that
$f_{\sigma }(E)=f_{0}(E;\mu_{\sigma},T_{\sigma})$, where
$f_{0}(E;\mu,T)=\{\exp [(E-\mu)/(k_{B}T)]+1\}^{-1}$ is the
Fermi-Dirac distribution function. $\mu_{\sigma}$ and $T_{\sigma}$
are determined by conservation of charge, spin and energy (see
Eqs.~\eqref{eq:steadystateequations}). The response matrix of the
spin valve
\begin{equation}%
\begin{pmatrix}
I\\
\dot{Q}%
\end{pmatrix}
=%
\begin{pmatrix}
G & GS\\
TGS & K
\end{pmatrix}%
\begin{pmatrix}
\Delta V\\
-\Delta T
\end{pmatrix}
\end{equation}
relates the charge and heat currents $I$ and $\dot{Q}$ to the biases
$\Delta V$ and $\Delta T,$ respectively. Below, we derive
expressions for the heat conductance $K$ and thermopower $S$, in the
presence of inter-spin energy exchange and for different magnetic
configurations.

The steady state potentials and temperatures can be determined from
Kirchhoff's laws for charge and energy for each spin.\cite{hatami}
For small $e\Delta V/k_{B},\Delta T\ll T_{\uparrow},T_{\downarrow}$,
\begin{align}
&  \sum_{i=L,R}I_{i,\sigma}+G_{\mathrm{sf}}(\mu_{\sigma}-\mu_{-{\sigma}%
})/e=0\label{eq:steadystateequations}\\
&  \sum_{i=L,R}\dot{Q}_{i,\sigma}+K^{\uparrow\downarrow}(T_{\sigma}%
-T_{-\sigma})+K_{\mathrm{e-ph}}(T_{\sigma}-T_{\mathrm{ph}})=0.\nonumber
\end{align}
\noindent Here
$I_{i,\sigma}=G_{i\sigma}(\mu_{\sigma}-\mu_{i})/e+G_{i\sigma
}S_{i\sigma}(T_{\sigma}-T_{i})$ is the charge current for spin
$\sigma$ through contact $i$,
$Q_{i,\sigma}=\mathcal{L}_{0}G_{i\sigma}T(T_{\sigma
}-T_{i})+G_{i\sigma}S_{i\sigma}T(\mu_{\sigma}-\mu_{i})/e$ is the
corresponding heat current, $G_{i\sigma}$ and $S_{i\sigma}$ are the
associated charge
conductances and Seebeck coefficients, and $\mathcal{L}_{0}=\pi^{2}k_{B}%
^{2}/(3e^{2})$ is the Lorenz number. Spin decay is described by the
(inter-)spin conductance $G_{\mathrm{sf}}=e^{2}\nu_{F}\Omega/\tau
_{\mathrm{sf}}$ for an island with volume $\Omega$, density of
states at the Fermi level $\nu_{F}$ and spin-flip relaxation time
$\tau_{\mathrm{sf}}$. The term $K_{\mathrm{e-ph}}$ describes the
interaction with the phonons at temperature $T_{\mathrm{ph}}$.
Inter-spin energy exchange is governed by the spin heat conductance
$K^{\uparrow\downarrow}=\mathcal{L}_{0}G_{\mathrm{sf}}T+K_{\mathrm{e-e}%
}^{\uparrow\downarrow}$, where the first term originates from the
spin-flip scattering and the second is due to e-e interactions. We
are allowed to discard the spatial dependence of the distribution
functions when the diffusion time $\tau _{D}=L^{2}/D$ in the island
with length $L$ and diffusion constant $D$ is shorter than both
$\tau_{\mathrm{sf}}$ and the spin thermalization time
$\tau_{\mathrm{st}}=\mathcal{L}_{0}e^{2}\nu_{F}T\Omega/(K_{\mathrm{e-ph}%
}+2K^{\uparrow\downarrow})$.

The in general lengthy solutions of
Eqs.~\eqref{eq:steadystateequations} are considerably simplified for
left-right symmetric conductances and Seebeck coefficients,
parameterized by $G_0=G_{\uparrow}+G_{\downarrow}$,
$P=(G_\uparrow-G_\downarrow)/G_0$, $S_0=(G_\uparrow
S_\uparrow+G_\downarrow S_\downarrow)/G_0$ and $P'=(G_\uparrow
S_\uparrow-G_\downarrow S_\downarrow)/(G_0 S_0)$ for both junctions.
In the antiparallel case the signs of $P$ and $P'$ in one of the
junctions are inverted.
In the parallel configuration the heat conductance becomes
\begin{equation}
K_{\mathrm{P}}=\mathcal{L}_{0}G_{P}T+\frac{2K_{\mathrm{e-ph}}r(1-P^{2}\gamma
)}{1-P^{2}\gamma+K_{\mathrm{e-ph}}/(\mathcal{L}_{0}G_{0}T)}%
\label{eq:pheatcond}%
\end{equation}
and in the antiparallel configuration it is
\begin{equation}
K_{\mathrm{AP}}=\mathcal{L}_{0}G_{P}T(1-P^{2}\gamma)+\frac{2K_{\mathrm{e-ph}%
}r}{1+K_{\mathrm{e-ph}}/(\mathcal{L}_{0}G_{0}T)}.\label{eq:apheatcond}%
\end{equation}
The factor $r=(T_{\mathrm{ph}}-T_{L})/(T_{R}-T_{L})-1/2$
parameterizes the phonon temperature on the island: If the phonons
are poorly coupled to the substrate, as for example in perpendicular
spin valves or in suspended structures,
$T_{\mathrm{ph}}=(T_\uparrow+T_\downarrow)/2$. For the P
configuration this yields $r=0$, whereas for the AP configuration we
get
$r=-K_{\uparrow\downarrow}P/[2(K_{\mathrm{e-ph}}+K_{\uparrow\downarrow}+\Lor
G_0 T)]$. In the opposite limit  $r=\pm1/2$, \textit{viz}.
$T_{\mathrm{ph}}$ is fixed to the bath temperature of the left or
right reservoir. The coefficient $\gamma
=[1+(K_{\mathrm{e-ph}}+2K^{\uparrow\downarrow})/(\mathcal{L}_{0}G_{0}T)]^{-1}$
describes inter-spin energy exchange. Factoring out the temperature
dependence of $K_{\mathrm{e-ph}}\propto T^{4}$ and
$K_{\mathrm{e-e}}^{\uparrow\downarrow }\propto T^{\nu+1}$ (see the
discussion below) yields $\gamma
=[1+(T/T_{\mathrm{ch,ph}})^{3}+(T/T_{\mathrm{ch,e-e}})^{\nu}+2G_{\mathrm{sf}%
}/G_{0}]^{-1}$, where the characteristic temperatures are $T_{\mathrm{ch,e-ph}%
}=[(\mathcal{L}_{0}G_{0}T^{4})/K_{\mathrm{e-ph}}]^{1/3}$, $T_{\mathrm{ch,e-e}%
}=[(\mathcal{L}_{0}G_{0}T^{\nu+1})/(2K_{\mathrm{e-e}}^{\uparrow\downarrow
})]^{1/\nu}$ for electron-phonon and electron-electron couplings,
respectively. The exponent $\nu$ depends on the dimensionality
($n$d) of the sample. We are here mainly interested in 3d samples
($\nu=3/2$) in which all sample dimensions exceed the thermal
coherence length $\xi_{T}=\sqrt{\hbar D/(2\pi k_B T)}$.

\begin{figure}[h]
\centering
\includegraphics[width=0.75\columnwidth]{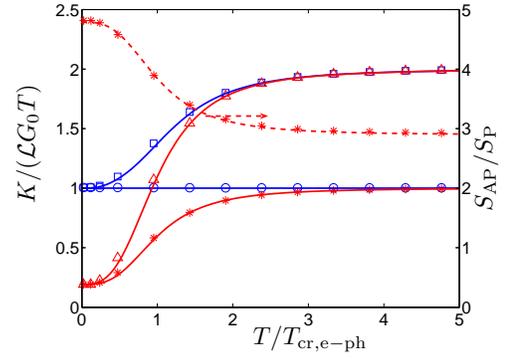}
\caption{(Color online): Temperature dependence of the heat
conductance $K$ (solid lines, left axis) and thermopower $S$ (dashed
line, right axis) of a structurally left-right symmetric spin valve
with $P=0.9$, $P^{\prime}=0.5$ and when the electron-phonon
relaxation dominates the inter-spin energy exchange. The lines are
plots of Eqs.~\eqref{eq:pheatcond}--\eqref{eq:apthermopower} and the
symbols have been calculated from the full nonequilibrium
distribution function, Eqs.~\eqref{eq:noneqdist} and
\eqref{eq:currents}. The results have been calculated for P
configuration with $r=0$ (circles) and $r=1/2$ (squares)
and AP configuration with $r=0$ (stars) and $r=1/2$ (triangles).}%
\label{fig:thermoelexamples}%
\end{figure}

In the parallel configuration the thermopower satisfies
$S_{\mathrm{P}}=S_{0}$ and in the antiparallel one\cite{nophononterm}
\begin{equation}
\frac{S_{\mathrm{AP}}}{S_{\mathrm{P}}}=\frac{1-PP^{\prime}+2G_{\mathrm{sf}%
}/G_{0}+\gamma P\left(  P-P^{\prime}-2P^{\prime}G_{\mathrm{sf}}/G_{0}\right)
}{1-P^{2}+2G_{\mathrm{sf}}/G_{0}}.\label{eq:apthermopower}%
\end{equation}

The temperature dependence of $K$ and $S$ is plotted in
Fig.~\ref{fig:thermoelexamples} for $K_{\mathrm{e-ph}}\gg K_{\mathrm{e-e}%
},\mathcal{L}_{0}G_{\mathrm{sf}}T$. For $T\ll \min(T_{\rm
ch,e-e},T_{\rm ch,e-ph})\equiv T_{\rm ch}$, the device operates as a
spin heat valve in which the heat current can be controlled by the
magnetization configuration. Contrary to the charge conductance,
however, the magnetoheat conductance
$(K_{\mathrm{P}}-K_{\mathrm{AP}})/K_{\mathrm{P}}$ vanishes for $T\gg
T_{\rm ch}$ or $\gamma \rightarrow 0$. Thus the presence of
inelastic scattering leads to a violation of the Wiedemann-Franz law
$K=\mathcal{L}_0 GT$ for $T \gtrsim T_{\rm ch}$.  The magnetothermopower $(S_{\mathrm{P}}-S_{\mathrm{AP}%
})/S_{\mathrm{P}}$ persists provided $P\neq P^{\prime}$.\cite{hatami} The measured heat conductance and thermopower as a
function of temperature and magnetic configuration may hence yield
unprecedented information on the energy relaxation in normal metals.

We now address the characteristic temperatures
$T_{\mathrm{ch,e-ph}}$ and $T_{\mathrm{ch,e-e}}$. The former can be
obtained directly from the Debye form for the heat conductance
between electrons and acoustic phonons,\cite{wellstood94}
$K_{\mathrm{e-ph}}=\frac{5}{2}\Sigma\Omega T^{4}$, valid for $T\ll
T_{\mathrm{Debye}}$. Here $\Sigma$ is the e-ph coupling constant\cite{giazotto06} and the factor $1/2$ takes into account spin
degeneracy. The characteristic temperature for electron-phonon
coupling thus reads
\begin{equation}
T_{\mathrm{ch,e-ph}}=\left(  \frac{\pi k_{B}^{2}}{15\hbar\Sigma\Omega}\right)
^{1/3}\left(  \frac{G_{0}h}{e^{2}}\right)  ^{1/3}. \label{eq:Tcreph}%
\end{equation}
For $T\gtrsim T_{\mathrm{Debye}}$, the electron -- acoustic phonon
scattering and thereby inter-spin energy exchange saturates. Optical
phonons start to contribute in this temperature regime, but are
disregarded here.

\begin{figure}[h]
\centering
\includegraphics[width=\columnwidth]{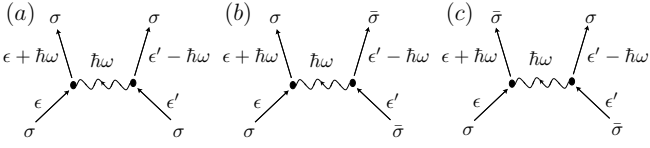}\caption{Electron-electron
scattering vertices. (a) Equal-spin scattering, which equilibrates the
electrons but does not thermalize the spins. (b) Spin conserving scattering
and (c) spin exchange scattering, which do thermalize the spins.}%
\label{fig:eescatterings}%
\end{figure}

The e-e scattering collision integrals with spin-dependent
distribution functions contain three terms 
\[
I_{\mathrm{e-e},\sigma}(\epsilon)=I_{\mathrm{(a)}}^{\sigma\sigma}%
(\epsilon)+I_{\mathrm{(b)}}^{\sigma,-{\sigma}}(\epsilon)+I_{\mathrm{(c)}%
}^{\sigma,-{\sigma}}(\epsilon),
\]
presented by the diagrams in Fig.~\ref{fig:eescatterings}.

Processes (b) and (c) induce inter-spin energy exchange, which can
be described in terms of a heat current flowing between two spin
ensembles,\cite{giazotto06}
\begin{equation}
\dot{Q}_{\mathrm{e-e}}^{\uparrow\downarrow}=\nu_{F}\Omega\int d\epsilon
\epsilon(I_{\mathrm{(b)}}^{\uparrow\downarrow}+I_{\mathrm{(c)}}^{\uparrow
\downarrow}).
\end{equation}
The direct spin current due to e-e interaction vanishes in the absence of
spin-orbit scattering, $\int d\epsilon(I_{\mathrm{(a)}}^{\uparrow\downarrow
}+I_{\mathrm{(b)}}^{\uparrow\downarrow}+I_{\mathrm{(c)}}^{\uparrow\downarrow
})=0.$ In 3d, to lowest order in spin particle and heat accumulation, $\left(
\mu_{\uparrow}-\mu_{\downarrow}\right)  /\left(  \mu_{\uparrow}+\mu
_{\downarrow}\right)  \ll1$ and $T_{s}/(T_{\uparrow}+T_{\downarrow})\ll1$, we
arrive at $\dot{Q}_{\mathrm{e-e}}^{\uparrow\downarrow}\approx(K_{(b)}%
^{\uparrow\downarrow}+K_{(c)}^{\uparrow\downarrow})(T_{\uparrow}%
-T_{\downarrow})$, where
\begin{subequations}
\begin{align}
K_{(b)}^{\uparrow\downarrow} &  =\frac{105\zeta(7/2)k_{B}^{7/2}T^{5/2}%
}{32[2\pi E_{T}(1+F)]^{3/2}\hbar},\label{eq:eeKb}\\
K_{(c)}^{\uparrow\downarrow} &  =\frac{FC}{(F+2)}\frac{[\pi^{2}\zeta
(3/2)+\frac{35}{16}\zeta(7/2)]k_{B}^{7/2}T^{5/2}}{2[2\pi E_{T}(F+1)]^{3/2}%
\hbar}.\label{eq:eeKc}%
\end{align}
\label{eq:eeK}
\end{subequations}
Here $C=-1+(F+1)^{3/2}$, $F>-1$ is the spin triplet Fermi liquid
parameter ($F=-1$ corresponds to the Stoner instability),
$E_{T}=\hbar D/\Omega^{2/3}$ is the Thouless energy proportional to
the inverse time it takes to diffuse over a length $\Omega^{1/3}$
and $\zeta(x)$ is the Zeta function. Summing the two contributions
from Eqs.~\eqref{eq:eeK} yields the characteristic temperature
\begin{align}
&  T_{\mathrm{ch,e-e}}=8\pi\left(  \frac{G_{0}h}{e^{2}}\right)  ^{2/3}%
\frac{E_{T}}{k_{B}}(F+1)\times\label{eq:Tcree}\\
&  \left\{ \frac{\pi(F+2)}{48FC\pi^{2}\zeta(3/2)+105[6+F(3+C)]\zeta
(7/2)}\right\}  ^{2/3}.\nonumber
\end{align}

In 1d and 2d structures the spin-flip contribution (c) has an
infrared divergence\cite{dimitrova07,chtchelkatchev08} that needs
to be regularized. As a result, the inter-spin energy exchange due
to e-e scattering becomes stronger and the corresponding
$T_{\mathrm{ch,e-e}}$ lower. This is especially relevant at low
temperatures and small structures since $\xi_T$ may exceed 100 nm at
$T\approx$ 1 K. We intend to analyze the resulting inter-spin energy
exchange in reduced dimensions in the future.

In order to assess the relevance of our results for realistic samples we
consider a disordered island of a spin valve coupled to the reservoirs via
tunnel contacts. For example, with $F=-0.3$ we get
\begin{align*}
T_{\mathrm{ch,e-e}}  &  \approx0.9\text{ }%
\operatorname{K}%
\times\frac{D}{0.001\text{ }%
\operatorname{m}%
^{2}/\text{s}}\left[  \frac{0.1\text{ }(%
\operatorname{\mu m}%
)^{3}}{\Omega}\frac{G_{0}}{0.01%
\operatorname{S}%
}\right]  ^{2/3}\\
T_{\mathrm{e-ph}}  &  \approx1\text{ }%
\operatorname{K}%
\times\left[  \frac{10^{9}\text{ }%
\operatorname{W}%
\operatorname{m}%
^{-3}%
\operatorname{K}%
^{-5}}{\Sigma}\frac{0.1\text{ }(%
\operatorname{\mu m}%
)^{3}}{\Omega}\frac{G_{0}}{0.01%
\operatorname{S}%
}\right]  ^{1/3}.
\end{align*}
Making the sample smaller and conductance larger increases both characteristic
temperatures, but the increase for $T_{\mathrm{ch,e-ph}}$ is slower. For
$\Omega=0.001$ ($%
\operatorname{\mu m}%
$)$^{3}$ and $G_{0}=1$ $%
\operatorname{S}%
$ we get $T_{\mathrm{ch,e-ph}}=22$ $%
\operatorname{K}%
$ whereas $T_{\mathrm{cr,e-e}}=400$ $%
\operatorname{K}%
$. We may therefore conclude that in spin valves with metallic
contacts and 3d spacers the inter-spin energy exchange due to e-e
interaction can be neglected. The spin thermalization rate with
$F=-0.3$ is
\begin{align*}
&  \frac{1}{\tau_{\mathrm{st}}}\approx\bigg[\frac{1}{20\mathrm{%
\operatorname{ns}%
}}\left(  \frac{T}{1%
\operatorname{K}%
}\right)  ^{3/2}\left(  \frac{0.001%
\operatorname{m}%
\mathrm{^{2}/s}}{D}\right)  ^{3/2}+\\
&  \frac{1}{25 \mathrm{%
\operatorname{ns}%
}}\left(  \frac{T}{1%
\operatorname{K}%
}\right)  ^{3}\left(  \frac{\Sigma}{10^{9}\text{ }%
\operatorname{W}%
\operatorname{m}%
^{-3}%
\operatorname{K}%
^{-5}}\right)  \bigg]\times\frac{10^{47}\text{ }%
\operatorname{J}%
^{-1}\text{ }%
\operatorname{m}%
^{-3}}{\nu_{F}}.
\end{align*}
The first term comes from e-e scattering and the second from e-ph
scattering. This rate exceeds the spin-flip
scattering rate $\sim10$ $%
\operatorname{GHz}%
$ at temperatures above $\sim10$ $%
\operatorname{K}%
$.

Above we assume that the electron energy distribution function is
well represented by Fermi-Dirac distributions with spin-dependent
chemical potentials and temperatures. This is not true in general,
since $f_{\sigma}(\epsilon)$ has the nonequilibrium form\cite{giazotto06,pothier97}
\begin{equation}
f_{\sigma}(\epsilon)=\frac{G_{L\sigma}f_{L}+G_{R\sigma}f_{R}+\nu_{F}%
e^{2}\Omega I_{\mathrm{coll}}[f_{\sigma},f_{-{\sigma}}]}{G_{L\sigma
}+G_{R\sigma}},\label{eq:noneqdist}%
\end{equation}
where $f_{L/R}=f_{0}(\epsilon;\mu_{L/R},T)$ are the distribution functions for
the reservoirs and $I_{\mathrm{coll}}$ describes all inelastic scattering
events. The charge ($n=0$) and heat ($n=1$) currents through contact $i$ then
become
\begin{equation}
I_{i}|\dot{Q}_{i}=\sum_{\sigma}\int d\epsilon(\epsilon
-\mu_{i})^{n}\frac{G_{i\sigma}}{e^{1+n}}(\epsilon)(f_{\sigma}(\epsilon)-f_{i}(\epsilon
)).\label{eq:currents}%
\end{equation}
Thermoelectric effects can be included by adding a weak energy dependence to
the conductances, $G_{i\sigma}(\epsilon)\approx G_{i\sigma}^{0}[1+c_{i\sigma
}(\epsilon-\mu_{i})]$, and expanding to linear order in $c_{i,\sigma}$.
Identifying $S_{i\sigma}=e\mathcal{L}_{0}c_{i\sigma}T$, we recover
Eqs.~\eqref{eq:apheatcond} and \eqref{eq:apthermopower} in the regime $e\Delta
V/k_{B},\Delta T\ll T_{L},T_{R}\approx T$ even in the absence of collisions
(\textit{i.e.}, $\gamma=1$). $\ $For $c_{i\sigma}=0$ and to linear order in
the applied bias, the nonequilibrium distribution \eqref{eq:noneqdist} is
identical to the quasiequilibrium one. Under these conditions, the collision
integrals can be calculated by replacing the full distribution functions by
the quasiequilibrium ones. Numerical solutions of the kinetic equations (see
Fig.~\ref{fig:thermoelexamples}) indicate that in linear response collisions
and finite $c_{i\sigma}$'s do not change this conclusion.

\begin{figure}[h]
\centering
\includegraphics[width=0.77\columnwidth]{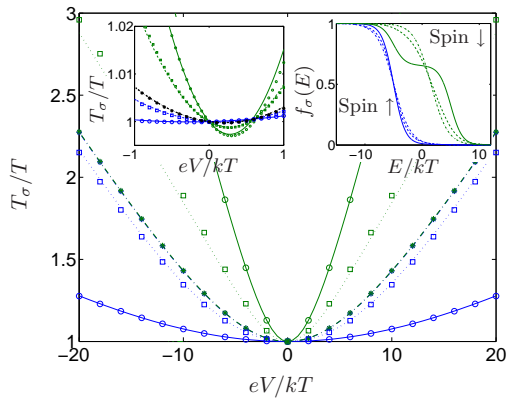}\caption{(Color
online): Spin-dependent effective temperature \textit{vs}. voltage
in an asymmetric spin valve with $P=0.9$, $P^{\prime}=0.5$ and
$G_{R}=0.1G_{L}$. The lines are calculated from
Eqs.~\eqref{eq:steadystateequations} and \eqref{eq:nonlinheatcur}
and the symbols from Eq.~\eqref{eq:efftemp} for numerical solutions
of the kinetic equations. The upper curves are for majority, the
lower for minority spins, and different strengths of e-e scattering
with $F=0$: no scattering (solid line and circles), weak scattering
with $E_{T}=0.05k_{B}T$ and $G_{L}=100e^{2}/h$  (dashed line and
squares), and strong scattering with $E_{T}=0.001k_{B}T$ and
$G_{L}=100e^{2}/h$  (dash-dotted line and stars). Here $T$ denotes
the temperature of the reservoirs. Left inset: behavior at low bias
with thermoelectric effects $c_{L}=10c_{R}=0.02/(kT)$. Right inset:
distribution function at $e\Delta V=10E_{T}$ with different strengths of e-e scattering.}%
\label{fig:efftemp}%
\end{figure}

Beyond linear response spin-dependent temperatures can strictly
speaking be invoked only in the presence of strong inelastic
scattering such that $T_{\uparrow}\approx T_{\downarrow}$.
Nevertheless we can define \textit{effective} electron temperatures
that satisfy the standard relation with the thermal energy density
in the Sommerfeld expansion:\cite{ashcroftmermin}
\begin{equation}
T_{\sigma}=\frac{\sqrt{6}}{\pi k_{B}}\sqrt{\int_{-\infty}^{\infty}[f_{\sigma
}(\epsilon)-1+\theta(\epsilon-\mu_{\sigma})]\epsilon d\epsilon}.
\label{eq:efftemp}%
\end{equation}
Proceeding with Fermi-Dirac distributions with effective spin-dependent
temperatures and chemical potentials, $\mu_{\sigma}$ and $T_{\sigma}$ can be
obtained from Eqs.~\eqref{eq:steadystateequations} by replacing the expression
for the charge and heat currents through contact $i$ with their nonlinear
counterparts,
\begin{align}
&  I_{\sigma}=\frac{G_{i,\sigma}}{e}\left\{  \mu_{\sigma}-\mu_{i}%
+\frac{c_{i\sigma}}{2}\left[  \mathcal{L}_{0}e^{2}(T_{\sigma}^{2}-T_{i}%
^{2})-(\mu_{\sigma}-\mu_{i})^{2}\right]  \right\} \nonumber\\
&  \dot{Q}_{i,\sigma}=G_{i,\sigma}\left[  \mathcal{L}_{0}(T_{\sigma}^{2}%
-T_{i}^{2})/2-(\mu_{\sigma}^{2}-\mu_{i}^{2})/(2e^{2})\right]
\label{eq:nonlinheatcur}\\
&  +G_{i\sigma}c_{i\sigma}(\mu_{\sigma}-\mu_{i})\left[  \mathcal{L}%
_{0}(T_{\sigma}^{2}+T_{i}^{2})/2-(\mu_{\sigma}^{2}-\mu_{i}^{2})/(6e^{2}%
)\right]  .\nonumber
\end{align}
These equations are obtained by a direct integration of
Eq.~\eqref{eq:currents} using Fermi-Dirac functions
$f_{i}(\epsilon)$ and $f_{\sigma}(\epsilon)$. We also have to
replace the linear-response forms of the spin mixing terms in
Eqs.~\eqref{eq:steadystateequations} by their forms far from
equilibrium. For example, for e-e scattering with $F=0$
we use $\dot{Q}^{\sigma\bar{\sigma}}=15\zeta(7/2)k_{B}^{7/2}(T_{\sigma}%
^{7/2}-T_{-{\sigma}}^{7/2})/[16\hbar(2\pi E_{T})^{3/2}]$.

In the absence of collisions and for weak thermoelectric effects it
can be proven by direct integration that the effective temperatures
defined by Eq.~\eqref{eq:efftemp} agree with those which follow from
heat conservation. In Fig.~\ref{fig:efftemp} we present a complete
numerical solution of the kinetic equations along with the results
from the quasiequilibrium heat balance equations from which we
conclude that the two approaches for calculating $T_{\sigma}$ agree
also in the presence of inter-spin energy exchange.

Spin heat accumulation cannot be directly measured by two-terminal
transport experiments in linear systems. In order to prove the
presence of a sizable $T_{s}$ far from equilibrium it should be
probed by spin-selective thermometry, such as a generalization of
the tunnel-spectroscopy in Ref.~\onlinecite{pothier97}, by measuring
the shot noise of the spin valve, or through electron spin
resonance.

In conclusion, we have shown that inter-spin energy exchange in a
spin valve affects the temperature and magnetic configuration
dependence of its thermoelectric properties. The different
thermalization mechanisms can be quantified by characteristic
temperatures, Eqs.~\eqref{eq:Tcreph} and \eqref{eq:Tcree}, above
which interaction effects become important. We introduce the concept
of spin heat accumulation via the spin-dependent effective electron
temperatures $T_{\sigma}$ in Fermi-Dirac distribution functions,
which can be used to describe transport properties beyond the linear
response regime. We demarcate the regime in which spin valves can be
employed to control heat currents. Other types of operations can be
envisaged as well, such as spin-selective cooling of the electrons
(see the left inset of Fig.~\ref{fig:efftemp}).

We thank P. Virtanen for discussions. This work was supported by the Academy
of Finland, the Finnish Cultural Foundation, and NanoNed, a nanotechnology
programme of the Dutch Ministry of Economic Affairs. TTH acknowledges the
hospitality of Delft University of Technology, where this work was initiated.


\begin{thebibliography}{99}
\bibitem{nophononterm}
The fact that $S$ is independent of $T_{\mathrm{ph}}$ can be
understood from the Onsager-Kelvin relation $\Pi=TGS$. The Peltier
coefficient $\Pi$ is measured without a temperature gradient and
therefore it cannot depend on $r$.
\end{thebibliography}
\end{document}